\documentclass[aps,prl,showpacs,twocolumn, longbibliography]{revtex4-2}

\usepackage{graphicx}
\usepackage{dcolumn}
\usepackage{bm}
\usepackage{color}
\usepackage{amsmath}
\usepackage{amssymb}
\usepackage{comment}
\usepackage{xcolor}

\usepackage[inline, final]{showlabels}
\renewcommand{\showlabelsetlabel}[1]
  {\hbox to 0pt{\hss\showlabelfont #1}}
\renewcommand{\showlabelfont}{\small\slshape\color{red}}

\begin{document}

\author{Wojciech Roga$^{1}$}
 \email{wojciech.roga@keio.jp}

\author{Baptiste Chevalier$^{1,2}$}
\email{baptiste.chevalier@etu.sorbonne-universite.fr}

\author{Masahiro Takeoka$^{1,3}$}
 \email{takeoka@elec.keio.ac.jp}

\affiliation{%
$^{1}$Department of Electronics and Electrical Engineering, Keio University, 3-14-1 Hiyoshi, Kohoku-ku, Yokohama 223-8522, Japan} 

\affiliation{
$^{2}$Sorbonne Université, CNRS, LIP6, 4 place Jussieu, F-75005 Paris, France
}

\affiliation{%
$^{3}$Advanced ICT Research Institute, National Institute of Information and Communications Technology (NICT), Koganei, Tokyo 184-8795, Japan
}

\date{\today}

\title{Fully Quantum Classifier}

\begin{abstract}In this paper we present a supervised machine learning quantum classifier. It consists of a quantum data re-uploading classifier with binary trainable parameters, the optimal values of which are found by a quantum search algorithm. We show that we can reach the quadratic speed-up in optimization trainable parameters compared to classical brute force search.
\end{abstract}

\maketitle

\section{Background and motivation}

In the modern data driven society, Artificial Intelligence and especially data classification are on the spotlight with a big impact on everyday life and economy. 

However the latter faces several challenges including the volume of data to process and the exploding number of parameters to train with direct impacts on security, privacy, and economy. That is why there is a need to always reduce computation time on one hand and to increase accuracy of the methods on the other.

In this way, the researchers have been exploring possibilities offered by quantum technology attracted by quantum features such as: exponential scaling of the dimensionality of quantum states with the number of elementary systems, parallelism of quantum superposition, non-classical correlations provided by quantum entanglement and computational speed-up offered by quantum algorithms \cite{cerezo_challenges_2022}.
Indeed the volume of research papers on the "quantum classifier" topic as well as the number of citations sharply increased within the last ten years, with no sign of saturation yet \cite{CitationReport}. 

Moreover, applications for machine learning have already been investigated through the use of Variational Quantum Algorithms exploiting nowadays NISQ devices \cite{cerezo_variational_2021}. 
Many competing models of quantum classifiers have been proposed for both fault-tolerant Quantum computers and NISQ devices as: QSVM \cite{rebentrost_quantum_2014}, Quantum decision trees, Quantum Nearest Neighbours, Quantum Kernel methods and many variational methods for QNN, QCNN \cite{cong_quantum_2019}, QTN \cite{huggins_towards_2019} (see for instance \cite{li_recent_2022} for a comprehensive review). Among them, one of the state of the art method is the so-called \textit{data re-uploading} \cite{perez-salinas_data_2020} that allows for universal classification. A recent research shows a notable interest in it with respect to other methods \cite{jerbi_quantum_2023}.

While concrete advantages presented by those methods are still discussed, as the training is still done in a classical fashion, it is clear that they suffer from the crucial problem of training a large number of parameters. If focusing on the case where parameters are binary, training the cost function with no further assumptions can turn out highly complex. Also there is no guarantees that the optimal configuration of parameters can be found because of the issues like barren plateaus, among others \cite{mcclean_barren_2018}. Alternative methods for hybrid algorithms have been proposed to bypass this problems, for example, iterative methods based on parameter-shifting instead of gradient-free optimizers \cite{mitarai_quantum_2018,schuld_evaluating_2019}. Other methods based on sequential optimization of parameters have been used too \cite{nakanishi_sequential_2020,roga_sequential_2023}.

Besides, some arguments suggest that quantum computers may help with the task of training many parameters due to an advantage of sampling over optimization in specific cases \cite{ma_sampling_2019,anari_sampling_2021} as well as amplitude amplification based algorithms \cite{brassard_quantum_2002, durr_quantum_1999,ambainis_quantum_2005}. Quantum training algorithms have already been investigated for a long time \cite{wiebe_quantum_2012,lloyd_quantum_2014,schuld_prediction_2016}. Although exponential speed-up is unachievable in most cases, except under very strict assumptions, quadratic speed-up can be reached in general. In \cite{schuld_supervised_2021} the author discuss how those methods could be applied in a Quantum Kernel setting. However, to our best knowledge, no one has yet considered applying a quantum training algorithm in recent QML methods as, for instance, data re-uploading. In this paper we present a fully-quantum classifier based on data re-uploading scheme. In addition to the universality of this method it benefits from a novel quantum training algorithm achieving a quadratic speedup over classical methods.

\section{Introduction}

\begin{figure}[h]
    \centering
    \includegraphics[scale=0.45]{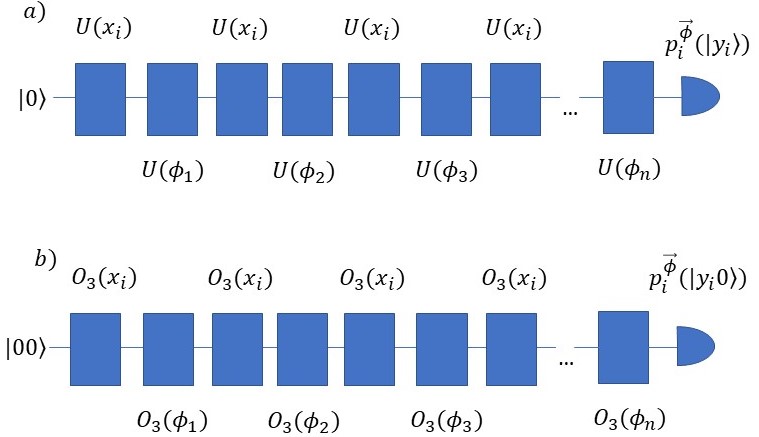}
    \caption{Scheme of data re-uploading circuit. $a)$ Training data $x_i$ associated with binary class label $y_i$ is introduced many times as a parameter of unitary transformations. Other unitary transformations contain trainable parameters used for training by optimization of the cost function fed by the output probability $p_i^{\vec \phi}(|y_i\rangle)$. The classification is based on the value of the probability $p_i^{\vec \phi}(|y_i\rangle)$ after training. $b)$ The same idea with two qubit circuits of only orthogonal transformations which we use in this paper as it is convenient to deal with real amplitude states at the output from the circuit.}
    \label{fig:citations}
\end{figure}

In this paper we present a supervised machine learning quantum classifier trained by a quantum algorithm. 
The starting point and elementary building block of our classifier is the single-qubit classifier with data re-uploading shown in figure \ref{fig:citations} $a)$ or its two qubit version shown in figure \ref{fig:citations} $b)$. The idea of the quantum classifier with data re-uploading was introduced in \cite{perez-salinas_data_2020} to show that even a single qubit system can compute complicated functions of parameters of the circuit. It consisted in a single qubit circuit with many one qubit gates such that the parameters of the gates can encode either the values of the coordinates of a training point or free trainable parameters. In a well trained classifier the thresholded probability of the chosen output state indicates the class. The technique of \cite{perez-salinas_data_2020} applied a supervised machine learning with hybrid quantum-classical training based on the optimization of the cost function fed by outputs from the one-qubit quantum circuit.

It was proven in \cite{perez-salinas_data_2020} that the counterpart of the universal approximation theory known from the theory of the single layer neural network holds in this case. In \cite{ono_demonstration_2023} the idea was extended to bosonic systems with even a single photon. The authors of the latter also provided the experimental proof of principle on the integrated photonics silicon chip. \\

For the algorithms we discuss in this paper it is convenient that the elementary classifiers produce only real amplitude output states. Therefore in this paper we replace the original idea of the one-qubit classifier with the SU(2) circuit shown in figure \ref{fig:citations} $a)$ by the two-qubit classifier with the circuit restricted to SO(3) transformations, as in figure \ref{fig:citations} $b)$. Due to the local isomorphism between SU(2) and SO(3) groups, the classifier with the 3D orthogonal transformations inherits the universality features of the single qubit classifier with the unitary circuit.  

\begin{figure}[h]
    \centering
    \includegraphics[scale=0.34]{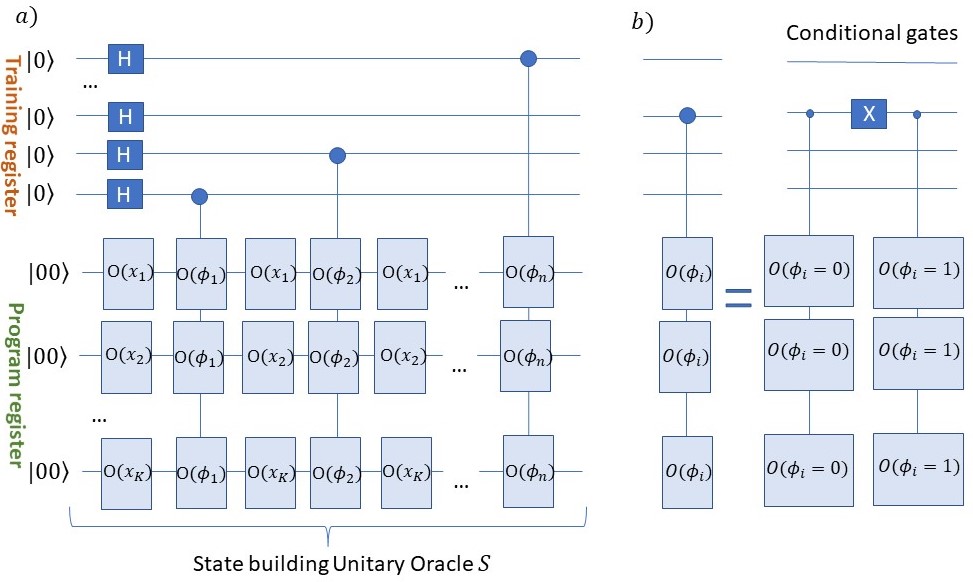}
    \caption{A parallelized version of the two qubit based quantum classifier with data re-uploading and orthogonal 3D transformations in which quantum search algorithms can be used for training. Parameters $x$ denote training data, parameters $\varphi$ adjustable parameters which we assume here to be binary.}
    \label{fig:fullyquantum}
\end{figure}

As the number of trainable parameters in the data re-uploading classifier can quickly grow the training of the circuit becomes challenging. Therefore in this paper, we explore the possibility of a parallel version of the classifier in which we introduce the dependence of the tunable parameters, which we assume now to be binary, on additional qubits. This scheme is shown in figure \ref{fig:fullyquantum}. 

Before proceeding let us comment on replacing the continuous angles with binary parameters. First of all, reducing continuous values to discrete does not remove the problem with the complexity of the search of the optimal set of angles on which we want to focus in this paper. Moreover, finding the best configuration of discrete angles is still interesting from the point of view of the search for the best starting point for the local search algorithms. Finally, composing rotations with discrete values, in principle, one can build arbitrary approximation of continuous rotations. Although we do not want to explore the last direction, this generalization is a straightforward extension of the proposed scheme.  

Conditioning the circuit parameters on additional qubits allows us to create a quantum superposition with the amplitudes related to different values of a specifically chosen objective function discussed in the next section. Following that, we propose to perform the quantum maximization algorithms \cite{durr_quantum_1999} and nonlinear transformations of complex amplitudes \cite{guo_nonlinear_2021} to maximize the objective function and train the classifier. Finally, we analyze the complexity of this scenario and observe that the fully quantum training can be advantageous with respect to classical maximum search algorithm with no assumptions achieving at most quadratic speed up. 

\section{Objective function and circuit description}

We assume that we have $k$ training points $\vec x=(x_1,...,x_k)$ each of which belongs to one of two classes $y_i\in \{0,1\}$. For each training point we prepare a copy of the circuit from figure \ref{fig:citations} $b)$ with the same $n$ trainable parameters $\vec\varphi=(\varphi_1,...,\varphi_n)$. As our objective function we take the product of the probabilities to measure the correct class $y_i$ for each data point in the first of two output qubits of elementary classifiers
\begin{equation}
P^{\vec y}(\vec\varphi)=p_{\vec\varphi,x_1}(|y_10\rangle)\ p_{\vec\varphi,x_2}(|y_20\rangle)\ ...\ p_{\vec\varphi,x_k}(|y_k0\rangle).
\label{pyf}
\end{equation}
This is the function of trainable parameters $\vec\varphi$ common to all copies of the circuit and training points $\vec{x}$. Considering here the probabilities of outputs from the elementary classifiers, fig. \ref{fig:citations} $b)$ with zero in the second qubit is our arbitrary choice and does not restrict generality. Objective function (\ref{pyf}) is closely related to the log likelihood. We want to train the circuits such that $P^{\vec y}(\vec\varphi)$ is maximized. When the training is finished we choose a threshold for the probability of output state $|10\rangle$ in a single copy of the elementary classifier from figure \ref{fig:citations} $b)$ with optimized $\vec\varphi$ and a testing point $x$. If the probability is above the threshold we classify $x$ as a point of class 1, otherwise we judge it as a point of class 0.

To utilize the quantum search algorithm we would like to have a superposition of all possible values of our objective function 
\begin{equation}
|\Psi\rangle=\sum_{\vec\varphi}\mathcal{A}^{\vec y}(\vec\varphi)|\vec \varphi\rangle,
\label{psi}
\end{equation}
where ${\mathcal{A}^{\vec y}(\vec\varphi)}^2=P^{\vec y}(\vec\varphi)$.
Indeed having state $|\Psi\rangle$ we would use D\"urr and Høyer maximization algorithm \cite{durr_quantum_1999} to find the maximum in, on average, $\mathcal{O}(2^{n/2})$ steps assuming access to appropriate oracles, the problem of which we will face in the next parts. This complexity would be advantageous with respect to a classical algorithm that without further assumptions would find $\max_{\vec \varphi} P^{\vec y}(\vec\varphi)$ in $\mathcal{O}(2^n)$ steps. This rough estimation motivates us to search for a circuit that outputs $|\Psi\rangle$. 

To construct $|\Psi\rangle$ we propose to use a circuit shown in figure \ref{fig:fullyquantum}. It consists of two registers, the "program register" with $2k$ qubits, where $k$ is the number of training points; and "training register" with $n$ qubits, where $n$ is the number of parameters in our elementary circuit. The elementary circuits with the same parameters but different training points repetitively re-uploaded are in each line of the program register. The qubits in the training register are prepared in the uniform superposition of all possible states. The orthogonal transformations in the program register with different $\vec \varphi$ are applied conditionally depending on the state of the training register. In consequence, the output state from this circuit is
\begin{equation}
|\Phi\rangle = \frac{1}{\sqrt{2^n}}\sum_{\vec y',\vec\varphi}\mathcal{A}^{\vec y'}(\vec\varphi)|\vec y'\rangle_{prog}|\vec \varphi\rangle_{train},
\label{phi}
\end{equation}
where the summation is over all possible configurations of $2k$-length bit strings for $\vec y'$ in the program register, and $n$-length strings for $\vec \varphi$ in the training register. This state reduces to (\ref{psi}) if the output of the program register $|\vec y'\rangle$ corresponds to the fixed program $\vec y$ there which describes classes of training points. So, we conclude that it is possible to conditionally construct the superposition of all objective functions (\ref{psi}) for given classification of training data which we call program. 

In what follows we will never perform selection of the program register and maximization over the training register separately. To find the optimal parameters $\vec \varphi$ for a given program $\vec y$ we will perform the single algorithm described as follows.

\section{Maximization algorithm}

In \cite{durr_quantum_1999} D\"urr and Høyer proposed a well-known algorithm that allows for finding maximum of a function $f$ with the domain of size $2^n$ in $\mathcal{O}(2^{n/2})$ steps on average. This algorithm assumes that one has access to the oracle $\Tilde{S}$ which prepares the state $\Tilde{S}|0\rangle=\sum_i f_i |i\rangle$ and the oracle that marks all entries $|i\rangle$ with $f_i$ larger than a chosen $f_j$. First, one samples a number $|k\rangle$ randomly $1\leq k\leq 2^n$. Then, using the Grover algorithm with the oracle that marks all entries $|j\rangle$ with $f_j>f_k$ one amplifies the probability of sampling from only the marked entries. This allows one to select an entry $|k'\rangle$ with $f_k'$ larger than $f_k$ with high probability. Repeating this procedure one can find the maximum with on average $\mathcal{O}(2^{n/2})$ applications of the oracle $\Tilde{S}$ and its inverse. The detailed analysis of the complexity of this protocol is given in \cite{durr_quantum_1999}, see also \cite{ambainis_quantum_2005}. 

Unfortunately this procedure cannot be directly applied in our case as we do not assume that we have access to the oracle $\Tilde{S}$ or the marking oracle from this algorithm. Instead, we have access to the unitary oracle $S$ that creates (\ref{phi}) the circuit of which is shown in figure \ref{fig:fullyquantum}.

In our optimization algorithm we propose to make use of this oracle in a protocol by Guo et al. \cite{guo_nonlinear_2021} that allows for nonlinear transformations of real and imaginary parts of the quantum amplitudes from (\ref{phi}). This algorithm is called nonlinear transformation of complex amplitudes $NTCA$. We want to use NTCA to suppress amplitudes smaller than the absolute value of the real amplitude $f_k$ of a randomly sampled entry $|k\rangle$ of the training register for a fixed program register. This allows us to follow the reasoning of D\"urr and Høyer \cite{durr_quantum_1999} described above replacing the marking oracle-based Grover algorithm with the NTCA. 

The NTCA algorithm \cite{guo_nonlinear_2021} assumes we have access to the unitary oracle $S$ that creates the $M$-qubit state $|\Phi\rangle$ the amplitudes of which we want to transform. In our case it will be the unitary transformation of the joint program and training registers of the circuit from figure \ref{fig:fullyquantum}, 
 hence $M=n+2k$. Authors of \cite{guo_nonlinear_2021} show how, using $S$ and its inverse four times, other single- and two-qubit operators, and 2 additional ancillary qubits, to build a Hermitian $2M+1$ qubit operator $X$ whose eigenvalues include the real parts $x_j$ and the imaginary parts $z_j$ of the amplitudes of $|\Phi\rangle$, i.e., $f_j=x_j+iz_j$. In our case, we need to deal only with real amplitudes.

Having Hermitian operator $X$, one can use an algorithm \cite{guo_nonlinear_2021} which depending on additional auxiliary qubits builds another Hermitian operator $Y$ whose eigenvalues are nonlinear functions $Q(x_j)$. 
The nonlinear functions that are allowed are approximated by degree-$d$ polynomials $Q'(x_j)$ 
and satisfy 
\begin{equation}
|Q(x)|\leq \frac{1}{4},\quad {
\rm for\ all}\ x\in[-1,1].
\label{pxleq}
\end{equation}

Process of building $Y$ is successful if the ancillary qubits measured give a desired output. The probability of the desired measurement of the ancillary qubits can be amplified by the amplitude amplification algorithm \cite{brassard_quantum_2002}. Altogether the total algorithm requires on average 
\begin{equation}
\mathcal{O}\left(d\gamma\sqrt{\frac{2^M}{\sum_i|Q(x_i)|^2}}\right)
\label{odgamma}
\end{equation}
applications of $S$ and its inverse, where $\gamma=\max_{x\in[-1,1]}\{|Q(x)|
\}$ which is a parameters that guarantees that (\ref{pxleq}) holds. Also a number $\mathcal{O}(dM)$ of elementary gates is required.

Next we prepare the state which is a uniform superposition of the eigenstates of operator $Y$, which are
\begin{equation}
W|k\rangle_{ad}|0\rangle_{da,B},
\label{Wk}
\end{equation} 
where $W$ is a circuit consisting of $S$ and its inversion on $2M+1$ qubit register and additional one and two qubit gates, see \cite{guo_nonlinear_2021}. This circuit acts on the so-called address register $|k\rangle$ and data register (in our case both consist of $M$ qubits), and auxiliary system $B$. 

The superposition of the eigenstates after acting on it by Hermitian operator $Y$ receives new amplitudes which are the eigenvalues of $Y$. Notice, what is important in our fully quantum classifier implementation of the algorithm \cite{guo_nonlinear_2021}, that as the input we can construct the superposition of states (\ref{Wk}) with fixed desired values of a part of the address register that corresponds to the fixed program in the program register. In consequence, at the output of the circuit we get the state which is 
\begin{equation}
|\Psi'\rangle=\sum_{\vec\varphi}Q\left(
\mathcal{A}^{\vec y}(\vec\varphi)
\right)
|\vec \varphi\rangle,
\end{equation}
where $\vec y$ is chosen by us. 

Our goal is to choose $Q$ 
as the effective activation function which in each iteration of the D\"urr and Høyer type maximization procedure suppresses the probabilities of the entries $\vec \varphi$ with the amplitudes which absolute values are smaller than a randomly selected reference. Then, by measuring the training register, we randomly select a new reference and repeat the procedure. We assume that for each reference $|\vec \varphi\rangle$ it is possible to efficiently calculate the amplitude classically which is needed for the appropriate choice of $Q$.

To do that one needs to compute classically the probabilities of the elementary classifiers from figure \ref{fig:citations} $b)$ for each training point with the specific parameters $\vec\varphi$ sampled previously. In order to estimate the cost of computing such amplitude, we consider the elementary quantum circuit made of $\mathcal{O}(n)$ two-qubit gates. Hence the full circuit can be represented as a single $4 \times 4$ unitary computed in time $\mathcal{O}(n)$. 
We need to repeat it $k$ times -- for each each training point. Since we already know the element $\vec y$ whose amplitude we want, it can be obtained in linear time from the $k$ individual amplitudes already computed for a total cost of $\mathcal{O}(nk)$. Thus the amplitude for a given $\vec \varphi$ and $\vec y$ can be calculated efficiently.

The number of iterations in D\"urr and Høyer algorithm is of the order of $n$. As the first random sampling gives us the value such that there are on average $2^n/2$ entries with larger amplitudes we sample from and each iteration reduces the sampling space on average by half. Notice that, since the sampling is done according to the state's distribution rather than uniformly, the space is actually reduced by more than half each time and the total number of iterations can be significantly smaller. Also because D\"urr and Høyer type algorithms keep reducing the size of the sampling space in every iteration, one is ensured, after iterating long enough, to find the global maximum that is achievable for this circuit architecture and for a given objective function (which can be different from the global maximum of the objective function in general).

\section{Complexity analysis and quantum speed-up}

The complexity of the classical brute force search of the set of binary parameters $\vec \varphi$ of length $n$ without further assumption is estimated as $\mathcal{O}(2^n)$.

The complexity of search algorithm presented in this manuscript is estimated as
\begin{equation}
\mathcal{O}\left(nd\gamma\sqrt{\frac{2^M}{\sum_i|Q(x_i)
|^2}}\right)
\end{equation}
where we must repeat each step of D\"ur and Høyer algorithm with suppression of small amplitudes by Guo  algorithm \cite{guo_nonlinear_2021} (\ref{odgamma}) in each of $\mathcal{O}(n)$ steps. Observing only dependence on $n$ is on average $\mathcal{O}(n\sqrt{2^{n+2k}})$ applications of the oracle circuit. The full circuit needed acts on $2(n+2k)+6$ qubits, where $k$ is the number of the training points. Also $\mathcal{O}(d(n+2k))$ elementary single and two qubit gates are needed in each step of the maximization algorithm. 

If $2k<n$, i.e., the number of training points is smaller than the number of parameters we start to observe the speed-up. Scaling with respect to $n$ when $k$ is kept fixed shows the quadratic speed up.

The proposed algorithm can be further improved if the maximum of the absolute value of the amplitudes in (\ref{psi}) could be estimated. Also knowing the program $\vec y$ one could try to design circuit with smaller dimension obtaining speed up by some factor not related to $n$.  

\section{Conclusions}

In this paper we propose a quantum circuit classifier with data re-uploading and parameters trained by a quantum algorithm. The algorithm is based on the D\"ur and Høyer maximization procedure \cite{durr_quantum_1999} with NTCA by \cite{guo_nonlinear_2021} replacing the Grover algorithm in the original version. The complexity of the algorithm is quadraticaly improved with respect to the classical brute force methods with no further assumptions. 

Notice that the same circuit as we propose here can be applied to many programs given by the vector of classes of the training points. 

\ \\

\textbf{\emph{Acknowledgements}}
This work was supported by JST Grant No. JPMJPF2221.

\bibliography{ref}

\end{document}